\input{aipcheck}

\documentclass[
    ,final            
  ]
  {aipproc}

\layoutstyle{6x9}

\begin{document}

\title{The dichotomy of the halo of the Milky Way}



\author{Daniela Carollo}{
  address={Research School of Astronomy and Astrophysics, Mount Stromlo Observatory, The Australian National University
  Mount Stromlo Observatory, Cotter Road, Weston, ACT 2611, Australia \& INAF, Osservatorio Astronomico di Torino, 10025, Pino
  Torinese, Italy},
  email={carollo@mso.anu.edu.au}
}

\author{Timothy C. Beers}{
  address={Department of Physics and Astronomy, Center for the Study of Cosmic Evolution, Joint Institute for Nuclear Astrophysics,
  Michigan State University, E. Lansing, Michigan 48824, USA},
  email={carollo@mso.anu.edu.au}
}

\begin{abstract}
We summarize evidence that the halo of the Milky Way comprises two
different, and broadly overlapping, stellar components. The two
structures exhibit different chemical compositions, spatial
distributions, and kinematics. These results were obtained through an
analysis of more than 20,000 calibration stars from the Sloan Digital
Sky Survey (SDSS). The duality of the stellar halo directly
impacts galaxy formation models, for the Milky Way and
other large spirals.

\end{abstract}

\maketitle


\section{Introduction}

The structure of the halo of the Milky Way has recently been revised
by the work of Carollo et al. (2007) \cite{Carollo07}. Confirming previous
speculations based on much smaller data sets, the halo is indeed
clearly divisible in two overlapping stellar components -- the inner
halo and the outer halo. The first structure dominates at R < 10-15
kpc, exhibits highly eccentric stellar orbits, is in slightly
prograde rotation, and comprises stars with a peak metallicity
around [Fe/H] $\sim -1.6$. The outer halo is dominant at R > 15-20
kpc, exhibits a much more uniform distribution in eccentricity,
includes stars on highly retrograde orbits, and possesses a peak
metallicity three times lower, i.e. [Fe/H] $\sim -2.2$. Previous
work provided hints that the halo of the Milky Way may not comprise
a single population, primarily based on analysis of the spatial
profiles (or inferred spatial profiles) for halo objects, and
possible indications of a net retrograde motion. In
any case, the past samples of tracer objects were not sufficiently
large to establish the dichotomy of the halo with confidence, and
usually were suitable only for consideration of a limited number of
the expected signatures of its presence. In Carollo et al. (2007) \cite{Carollo07}
all of the expected signals for the presence of two different
stellar halo populations (different spatial distributions,
kinematics, and chemical compositions) are now seen. This was made
possible through the analysis of a homogeneously selected sample of
more than 20,000 stars, originally obtained as calibration data
during the course of the Sloan Digital Sky Survey (SDSS, \cite{York00}).

\section{Derivation of the kinematic and orbital parameters}

The total number of unique stars in the sample is 20,366, and
comprises mainly F and G main-sequence turnoff stars. The apparent
magnitude range is 15.5 < g$_{0}$ < 17.0 (spectrophotometric
calibration stars), and 17.0 < g$_{0}$ < 18.5 (telluric calibration
stars). The color ranges are 0.6 < (u-g)$_{0}$ < 1.2 ; 0 < (
g-r)$_{0}$ < 0.6 (see Figure 1).

\begin{figure}[htp]
\includegraphics[width=7.6 cm]{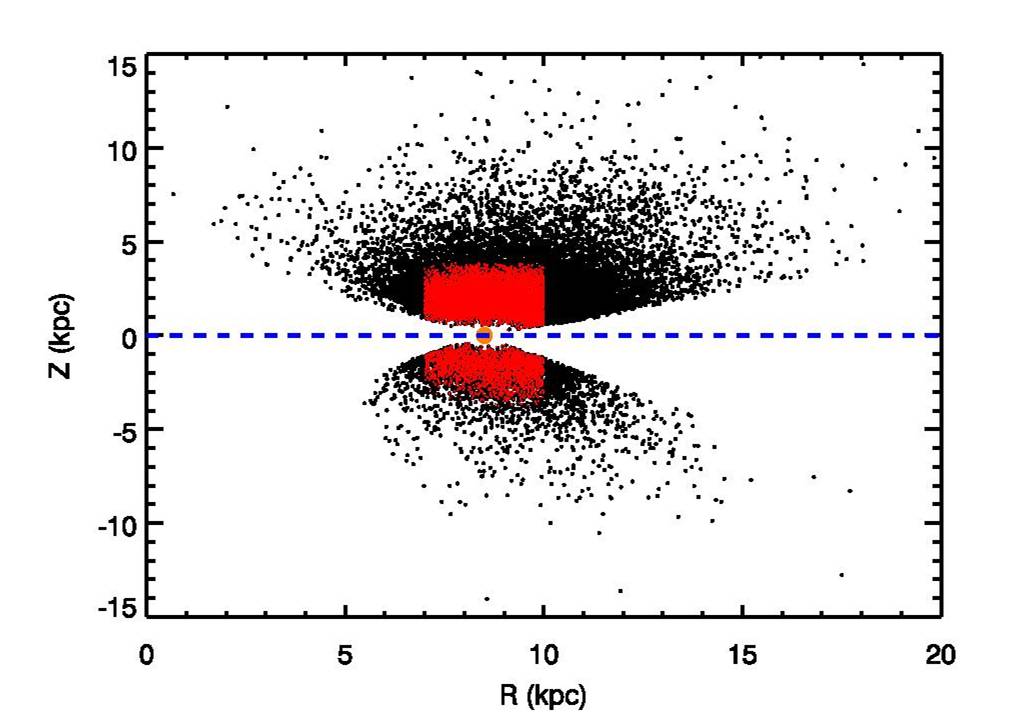}
\caption{The spatial distribution of the SDSS-DR5 calibration stars
in the Z-R plane, where Z is the distance from the Galactic plane,
and R is the distance from the center of the Galaxy projected onto
the Galactic plane. The dashed blue line represents the Galactic
plane, while the filled orange dot is the position of the Sun, at  Z
= 0 kpc and R = 8.5 kpc.  The ``wedge shape'' of the selection area
is the result of limits of the SDSS footprint in Galactic latitude.
The red points indicate the 11,458 stars that satisfy our criteria
for a ``local sample'' of stars, having  7 kpc $<$ R $<$ 10 kpc, with
distance estimates from the Sun $<$ 4 kpc, and with viable
measurements of stellar parameters and proper motions.}
\end{figure}

In order to derive the full space motions of the stars, we require astrometry
(positions and proper motions), radial velocities, and distances. The SDSS
provides positions with an accuracy of 0.1$"$, while proper motions are provided
by the re-calibrated USNO-B2 catalogue (\cite{Munn04}), with an accuracy of 3-4
mas/yr. The radial velocities are derived from matches to an external library of
high-resolution spectral templates with accurately known velocities; its
accuracy is around 5-20 km/s, depending on the S/N of the spectrum. The
distances are evaluated using the cluster fiducials of Beers et al. (2000)
\cite{Beers00}; the accuracy is around 10-20\%. The SEGUE Stellar Parameter
Pipeline (SSPP, \cite{Lee08a, Lee08b,Allende08}) provides the stellar physical
parameters, i.e. effective temperature, surface gravity, and metallicity, with
accuracies of 100~K, 0.25 dex, and 0.20 dex, respectively.

In order to obtain the best available estimates of the kinematic and orbital
parameters for the stars in our sample, we consider only those stars satisfying
several cuts: (1) A selection in the effective temperature range 5000~K <
T$_{\rm eff}$ < 6800~K, over which the SSPP is expected to provide the highest
accuracy, reducing the number of stars to 19,687, (2) A selection for stars in
the sample with distances d $<$ 4 kpc from the Sun, in order to restrict the
kinematical and orbital analyses to a local volume (where the assumptions going
into their calculation are best satisfied), reducing the number of stars to
15,435. The choice of stars in a local volume reduces errors in the derived
transverse velocities, which scale with distance from the Sun. The number of
stars in the remaining sample is 11,458. The proper motions, used in combination
with radial velocities and the estimated distances, provide the information
required to calculate the full space motions (U,V,W) of the stars relative to
the Local Standard of Rest (LSR). We have also obtained the velocity components
of the stars in a cylindrical reference frame with origin at the Galactic center
(V$_{R}$, V$_{\Phi}$, V$_{Z}$).

The orbital parameters are derived adopting a St\"{a}ckel-type gravitational
potential that comprises a flattened oblate disk and a spherical massive dark
halo (\cite{deZeeuw85,Dejonghe88}). The quantities obtained are
r$_{peri}$, which correspond to the closest approach of an orbit to the Galactic
center, and r$_{apo}$, which is the farthest extent of an orbit from the
Galactic center. The orbital eccentricities and the maximum distance of stellar
orbits above or below the Galactic plane, Z$_{max}$, are also evaluated.

\begin{figure}[htp]
\centering
\includegraphics[height=9 cm,width=7 cm]{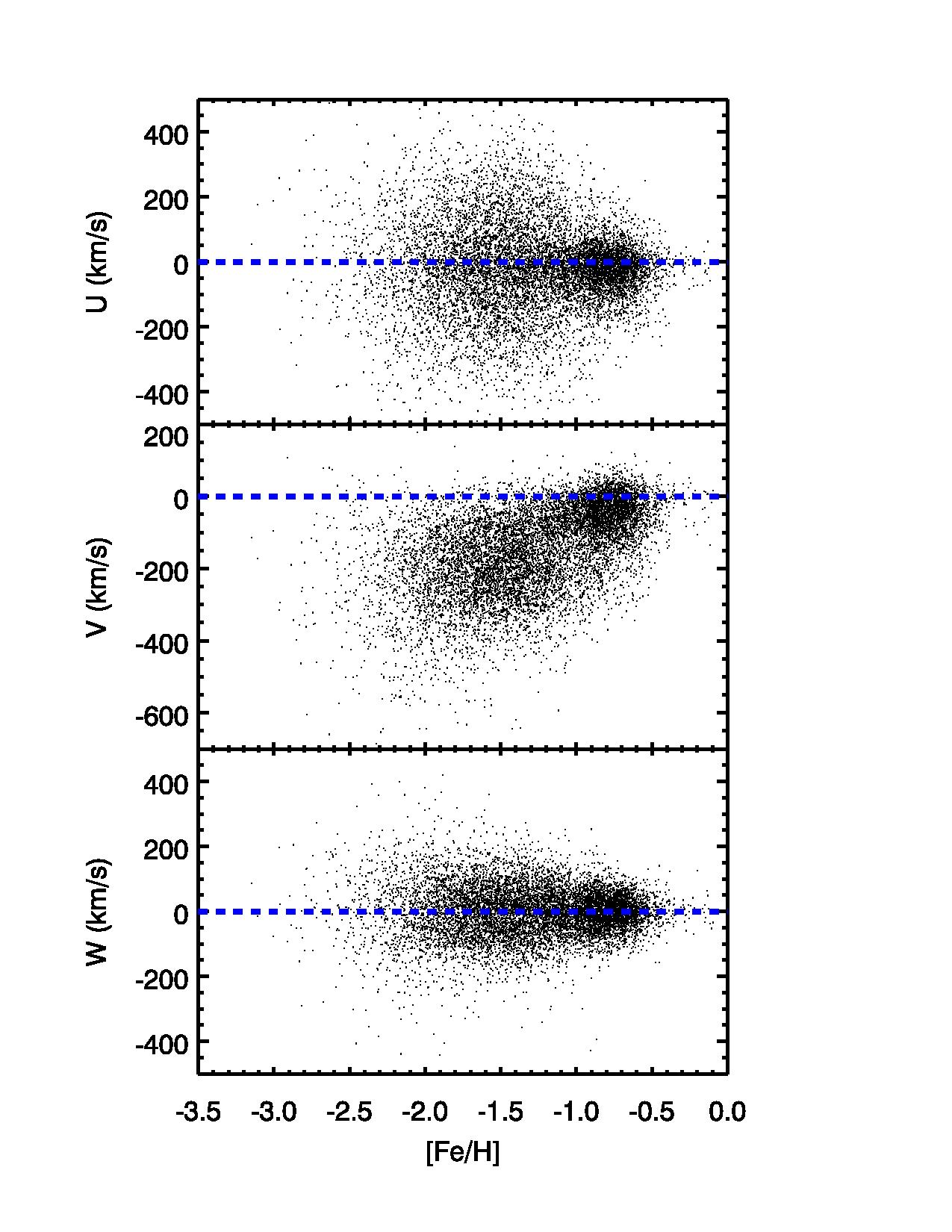}

\caption{The U, V, and W components of the space motions of stars in
our sample as a function of [Fe/H].  A blue dashed line at 0 km/s
has been added in each panel for reference. In the metallicity range
$-1.2 <$ [Fe/H] $< -0.3$, all three panels exhibit a low-velocity
dispersion population of stars with mean values of the velocity
components near 0 km/s, and dispersion in velocities of around 40-50
km/s.  These are the stars of the thick-disk and metal-weak
thick-disk populations. At metallicities below [Fe/H] $\sim -1.2$,
the large velocity dispersions (on the order of 100 to 150 km/s)
observed in each component are associated with the broadly
overlapping inner- and outer-halo populations of stars.}
\end{figure}


\section{Evidence for the dichotomy of the galactic halo}

\begin{figure}[t]
\centering
\includegraphics[angle=90, height=12 cm,width=14 cm]{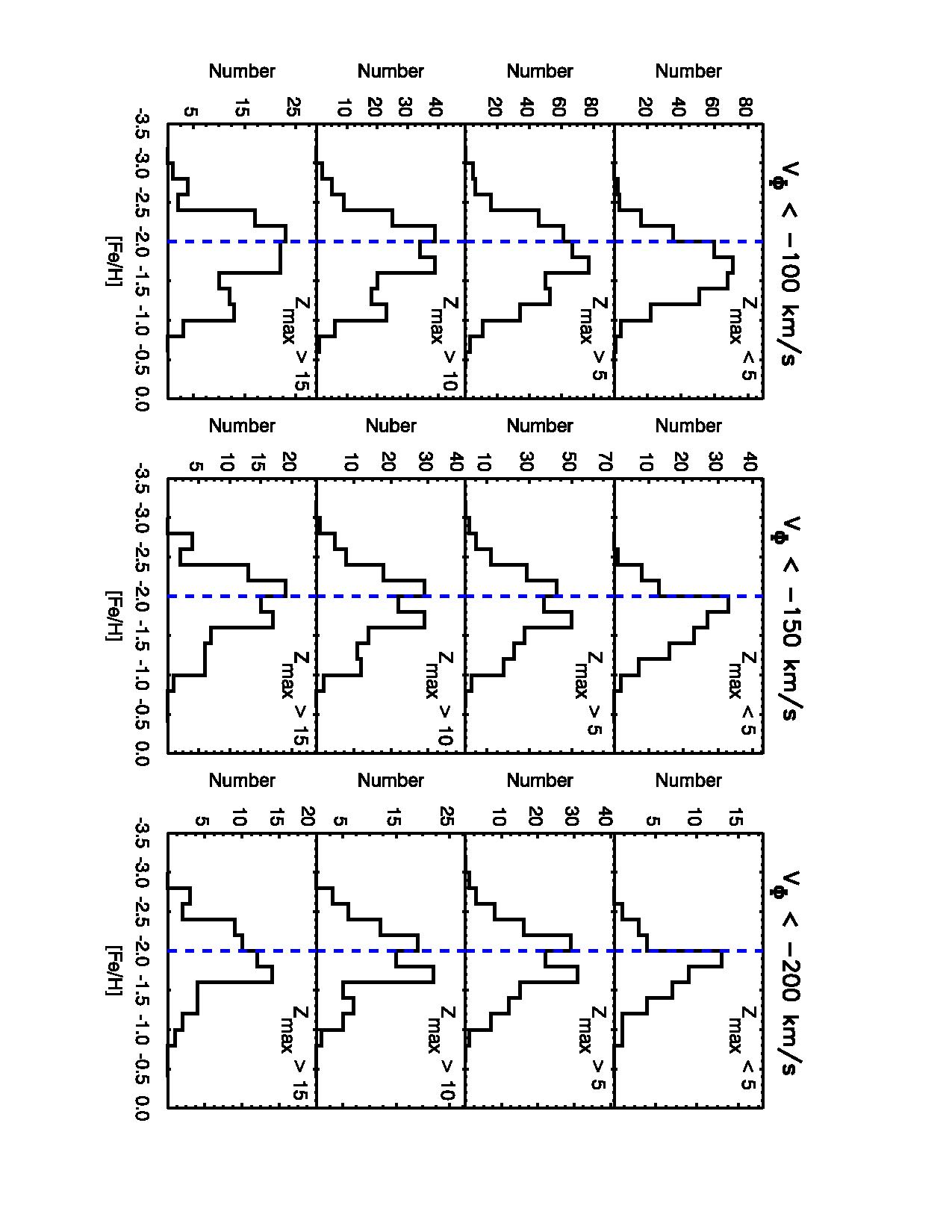}

\caption{The distribution of [Fe/H]  for the stars in our sample on
highly retrograde orbits.  The panels show various cuts in
V$_{\Phi}$, and for different ranges of Z$_{max}$. A blue dashed
line at [Fe/H] $= -2.0$ is added for reference. The left-hand column
applies for stars with V$_{\Phi} < -100$ km/s; the clearly skewed
distribution of [Fe/H] exhibits an increased contribution from lower
metallicity stars as one progresses from the low (Z$_{max} <$ 5 kpc)
to the high (Z$_{max} >$ 15 kpc) subsamples. Simultaneously, the
predominance of stars from the inner-halo population, with peak
metallicity at [Fe/H] $\sim -1.6$, decreases in relative strength,
and shifts to lower [Fe/H]. Similar behaviors are seen in the
middle and right-hand columns, which correspond to cuts on V$_{\Phi}
< -150$ km/s and $-200$ km/s, respectively.}
\end{figure}

There exist multiple signatures for the dichotomy of the Galactic halo; space
precludes a discussion of all of them (see \cite{Carollo07}). The change in
the distribution of stellar metallicity with the rotational velocity, V or
V$_{\Phi}$, and Z$_{max}$ is one of the stronger pieces of evidence. Figure 3
shows the Metallicity Distribution Function (MDF), for different values of
Z$_{max}$, for stars with retrograde orbits. Stars from the disk population,
which possess highly prograde orbits, cannot be present in this plot. We
performed a Kolmogorov-Smirnoff test of the null hypothesis that the MDFs of
stars shown in the lower panels for the individual cuts on V$_{\Phi}$ could be
drawn from the MDFs of the same parent population as those shown in the upper
panels; the null hypothesis is strongly rejected at a high level of statistical
significance. We thus conclude that the halo comprises stars with intrinsically
different distributions in metallicity. This behavior, as well as an imbalance
of the energy distributions for these stars (see \cite{Carollo07} for
details), would not be expected if ``the halo'' is a single entity. The local
sample of SDSS calibration stars exhibits a (highly statistically significant)
net retrograde rotation for the outer-halo component of V$_{\Phi}$ $\sim -70$
km/s. This value is in agreement with the Frenk \& White analysis
(\cite{Frenk80}), in which the rotational velocity is estimated on the basis of the
radial velocity and the distance alone, so that there is no propagation of
errors on the proper motions into the calculation of the kinematical quantities.

Other evidence for the duality of the Galactic halo can be obtained from an
inversion of the orbital properties of the sample to obtain relative densities
of the tracer population as a function of distance. Inspection of Figure 4
reveals that the inner regions are rather flattened at higher [Fe/H]. In
particular, for the metallicity bin $-1.4 <$ [Fe/H] $< -1.8$ (third and fourth
panel), the axis ratio is around 0.6. As [Fe/H] sweeps toward lower values, the
spatial distribution becomes less flattened, and then roughly spherical at
[Fe/H] $< -2.2$ (axis ratio $\sim$ 0.9). This indicates that the inner-halo
population dominates locally for stars with [Fe/H] $> -2$, whereas the
outer-halo population dominates for all [Fe/H] at distances R $>$ 15-20 kpc, and
also locally for stars with [Fe/H] $< -2$. Carollo et al. (2007) show that the
the inner and outer halos also exhibit different orbital characteristics as a
function of metallicity.

\begin{figure}[t]
\centering
\includegraphics[angle=-90,width=.7\linewidth]{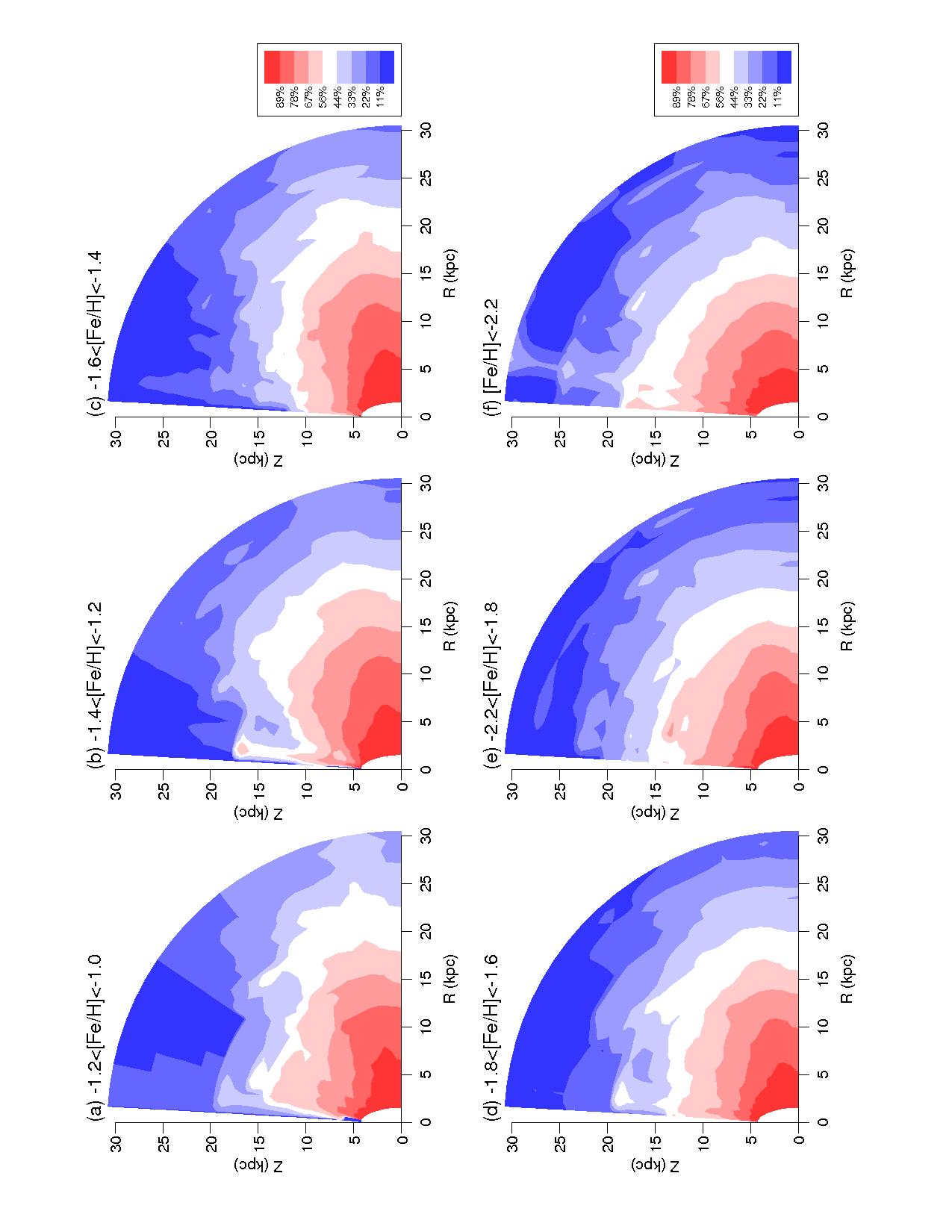}
\caption{Equidensity contours of the reconstructed global density
distributions for stars in our sample with various metallicities.
The global density distributions are constructed from the sum of the
probability density of an orbit at each location in this plane, with
a weighting factor being inversely proportional to the corresponding
density at the currently observed position of the star.}
\end{figure}

\section{Implications of the dichotomy of the halo}

The global properties of the outer-halo population clearly indicates that its
formation is distinct from both the inner-halo population and the disk
components of the Milky Way. Within the context of the $\Lambda$CDM paradigm, a
possible scenario of the formation of the outer halo is the dissipationless
chaotic merging of smaller subsystems within a pre-existing
dark-matter-dominated halo. These subsystems were subjected to tidal disruption
in the outer part of the dark-matter halo due to their lower masses
(\cite{Bekki01}). As candidate (surviving) counterparts for such subsystems, one
might consider the currently observed low-luminosity dwarf spheroidal galaxies
surrounding the Galaxy, in particular the most extreme cases recently identified
from the SDSS (\cite{Belokurov06,Belokurov07}).

The difference in the MDF of the inner/outer halo components has the important
consequence that the lowest metallicity stars may be associated with the outer
halo. This information could be used to search for the most metal-poor stars in future
surveys, by selection of stars with highly retrograde proper motions. Other
important issues related to the chemical difference of the two halo components
should also be explored, e.g., the determination of the primordial
lithium abundance from observations of the most metal-poor stars
(\cite{Bonifacio07}), and the existence of a possible relationship between the numbers of
carbon-enhanced metal-poor stars as a function of declining metallicity
(\cite{Lucatello06}), and as a function of distance from the Galactic plane
(\cite{Tumlinson07}).

Very recently, Koch et al. (2007) \cite{Koch07} reported the existence of a strong
gradient in the abundance distribution of stars located along M31's minor axis,
and towards their outer halo fields, which exhibit a metallicity [Fe/H] $< -2$ .
This finding is consistent with the present results found for the Milky Way.
Although they may differ in detail, it appears that the two dominant galaxies of
the Local Group have undergone similar accretion histories.

\begin{theacknowledgments}

The authors are grateful to the organizers for providing assistance with travel
and accomodation expenses.  This work also received support from grants AST
07-07776 and PHY 02-15783; Physics Frontier Center / Joint Institute for
Nuclear Astrophysics (JINA), awarded by the US National Science Foundation.

\end{theacknowledgments}


\bibliographystyle{aipprocl} 

\end{document}